%% file: apssamp.tex
\pdfoutput=1
\documentclass[aps,prd,amsmath,floats,floatfix, twocolumn,
superscriptaddress,nofootinbib,showpacs,longbibliography]{revtex4-1}
\input{preamble}
\usepackage{physics}
\usepackage{orcidlink}
\usepackage{multirow}
\usepackage{enumitem}
\DeclareMathAlphabet{\mathpzc}{OT1}{pzc}{m}{it}

\begin{document}

\preprint{APS/123-QED}

\title{A pipeline to search for signatures of line-of-sight acceleration in gravitational wave signals produced by compact binary coalescences}

\newcommand{\IUCAA}{\affiliation{Inter-University Centre for Astronomy and Astrophysics, Post Bag 4, Ganeshkhind, Pune 411007, India}}
\newcommand{\CITA}{\affiliation{Canadian Institute for Theoretical Astrophysics, University of Toronto, 60 St. George Street, Toronto, ON M5S 3H8, Canada}}
\newcommand{\AEI}{\affiliation{Max Planck Institute for Gravitational Physics (Albert Einstein Institute), D-30167 Hannover, Germany}}
\newcommand{\LU}{\affiliation{Leibniz University Hannover, D-30167 Hannover, Germany} }
\newcommand{\UiS}{\affiliation{Department of Mathematics and Physics, University of Stavanger, NO-4036 Stavanger, Norway} }

\author{Avinash Tiwari\,\orcidlink{0000-0001-7197-8899}}
\email{avinash.tiwari@iucaa.in}
\IUCAA
\author{Aditya Vijaykumar\,\orcidlink{0000-0002-4103-0666}}
\email{aditya@utoronto.ca}
\CITA
\author{Shasvath J. Kapadia\,\orcidlink{0000-0001-5318-1253}}
\email{shasvath.kapadia@iucaa.in}
\IUCAA
\author{Shrobana Ghosh\,\orcidlink{0000-0002-7654-478X}}
\AEI
\LU
\author{Alex B. Nielsen\,\orcidlink{0000-0001-8694-4026}}
\UiS

\hypersetup{pdfauthor={Tiwari et al.}}

\date{\today} 

\begin{abstract}
\noindent Compact binary coalescences (CBCs), such as merging binary black holes (BBHs), binary neutron stars (BNSs), or neutron star black holes (NSBHs), hosted by dense stellar environments, could produce gravitational waves (GWs) that contain signatures of line-of-sight acceleration (LOSA) imparted by the environment's gravitational potential. We calculate the Post-Newtonian (PN) corrections to the $(2,\,2)$ mode GW phase due to a finite LOSA, starting from the leading order at -4 PN below the quadrupole order, up to 3.5 PN above the leading order correction. We do so for binaries whose component spins are aligned with the orbital angular momentum, as well as for binaries with non-zero tidal deformation. We implement these corrections into the LIGO-Virgo-Kagra (LVK) collaboration's flagship parameter estimation (PE) software \textsc{Bilby\_tgr}. We study the systematics associated with recovering LOSAs. We find that, when the injection and recovery waveform models are identical, LOSAs are recovered as expected. We test the robustness of the pipeline against waveforms with strong higher-mode signatures or signatures of beyond-general-relativistic (beyond-GR) effects, to delimit the range of applicability of our GR-consistent quasi-circular LOSA-corrected waveforms. 
\end{abstract}

\maketitle

\section{Introduction}
\noindent The LIGO-Virgo-KAGRA (LVK) network of gravitational-wave (GW) detectors~\cite{TheLIGOScientific:2014jea, TheVirgo:2014hva, KAGRA:2020tym} has observed $\sim 90$ stellar-mass compact binary coalescence (CBC) events in its first three observing runs (O1, O2, O3)~\cite{LIGOScientific:2018mvr, LIGOScientific:2020ibl, LIGOScientific:2021usb, KAGRA:2021vkt, nitz_ogc3, ogc4_Nitz_2023, venumadhav_2022_o3a, wadekar2023newblackholemergers, mehta2024newbinaryblackhole}, and will likely more than triple this number by the end of the ongoing O4 \cite{2020LRR....23....3A}. The vast majority of these are binary black hole (BBH) mergers, although binary neutron star (BNS)~\cite{TheLIGOScientific:2017qsa, Abbott:2020uma} and neutron star black hole (NSBH)~\cite{LIGOScientific:2021qlt} mergers have also been observed. Probing and constraining the environments that host these CBCs is an area of active research \cite{Mapelli:2020vfa}. Nevertheless, confidently ascertaining a CBC's host environment is challenging. This, in part, is because there is increasing evidence to suggest that no single formation channel can explain the spectra of parameters of the CBCs observed~\cite{Zevin:2020gbd}. Perhaps the only way to identify a host environment is if it imprints itself on the CBC's GW waveform, or if the CBC emits an electromagnetic (EM) counterpart.

Past work \cite{Yunes2011, Bonvin2017} has shown that a finite CBC centre of mass (CoM) line-of-sight acceleration (LOSA) (and its higher time derivatives)  would modulate the GWs with respect to those emitted by non-accelerating CBCs. The kinematic parameters could then, in principle, be extracted, and the environment's gravitational potential reconstructed \cite{Tiwari:2024pvb}. Thus, accelerated motion could be one means to conclusively determine a CBC's host environment on a single event basis. It is therefore of considerable interest to search for signatures of accelerated motion in LVK's CBC detections. 

Even if a LOSA measurement is not complemented by measurements of higher time derivatives, it could still shed some light on the nature of the environment. In a different context, this problem can be posed as a consistency test of the observed signal with GR --- a detectably accelerating source would be inconsistent with GR.  
Therefore, in this work, we focus on how a constant acceleration of the CoM modulates the CBC waveform and develop a pipeline to search for (and constrain) LOSA in confirmed GW detections. 

A CBC moving with a constant relative velocity with respect to a GW detector is indistinguishable from a stationary CBC with rescaled masses---a byproduct of the well-known mass-redshift degeneracy. In other words, the Doppler shift caused by the projection of the velocity along the line-of-sight is completely degenerate with the chirp mass of the CBC.
However, a CBC whose CoM has a finite LOSA produces a time-varying Doppler shift, which effectively breaks the chirp-mass redshift degeneracy, 
and leads to modulations in the GW phase, starting at $-4{\rm PN}$ \cite{Bonvin2017, Vijaykumar_2023} (leading order). LOSA is therefore a low-frequency effect, readily observable by detectors sensitive in the low-frequency regime \footnote{See \cite{Vijaykumar_2023} for LOSA detectability in various observing scenarios involving ground and space-based detectors, and \cite{Tiwari:2023cpa} for prospects of LOSA detections with space detectors for CBCs in globular clusters.}. Nevertheless, a sufficiently large LOSA could also potentially be detectable with the LVK network at design sensitivity. In \cite{Vijaykumar_2023}, we derived the phase correction to the dominant $22$ mode phase for a non-spinning point-particle CBC containing up to $3.5 {\rm PN}$ higher order terms.
In this work, we consider aligned-spin CBCs, as well as those with finite tidal deformability. We provide the corresponding LOSA PN corrections to 3.5 PN order in this paper, and an accompanying \textsc{LOSA-Pipe} software pipeline integrated with the parameter estimation software \textsc{Bilby\_TGR}~\cite{2024zndo..10940210A}. 

While our work provides LOSA phase corrections to the dominant 22 mode of the GW waveform (Section~\ref{sec:phase_corr_der}), it neglects special relativistic effects pertaining to the motion of the CoM. Accordingly, we map out the domain of validity for which our LOSA-PN-corrections may be reliably used (Section~\ref{sec:approx_validity}). We additionally test for systematics associated with LOSA-corrected CBC waveforms. In particular, we first ascertain that when the injected and recovered LOSA-corrected waveform models are identical, all parameters of the waveform are recovered consistently (Section~\ref{sec:LOSA_PE}). We then list a series of potential LOSA mimickers, where non-inclusion of certain physical or beyond-GR effects in a non-accelerated GW waveform could produce spurious values of LOSA when recovered with a standard LOSA-corrected (`for the 22-mode only) CBC (Section~\ref{sec:pot_sos}). In this context, we also demonstrate how using 22-mode only LOSA correction while recovering a non-accelerating GW190814-like CBC could lead to spurious LOSA recoveries (Section~\ref{sec:GW190814_like_analysis}). We accordingly point out to the reader that the results of \cite{2024arXiv240101743H} should be interpreted with caution. We end the paper by listing our selection criteria pertaining to CBCs for which our LOSA PN corrections may be reliably used to probe accelerated CoM motion (Section~\ref{sec:sel-crit}), and summarizing our work (Section~\ref{sec:conclusion}). 

\section{PN correction to the GW phase due to finite LOSA}\label{sec:phase_corr_der}
Let $M$ and $M_{\rm LOSA}$ be the detector frame total masses --- so that they already account for cosmological redshift and the Doppler shift due to any constant peculiar velocity --- of a non-accelerating and an accelerating CBC, respectively, and $\Gamma_1 \equiv a/c$ be the observed LOSA of the center of mass (CoM) of the accelerating CBC. Then, assuming that $| \Gamma_1 (t_{\rm o} - t_{\rm c}) | \ll 1$, we can write:
\begin{equation}
    \label{eq: redshifted_mass}
    M_{\rm LOSA} = M(1 + \Gamma_1 (t_{\rm o} - t_{\rm c}))
\end{equation}
where $t_{\rm o}$ is the observation time and $t_{\rm c}$ is the time at coalescence. 

Under the stationary phase approximation, we can write the modulated inspiral GW waveform as \cite{PhysRevD.80.084043}:
\begin{equation}
    \label{eq: WF}
    \tilde{h}(f) = \mathcal{A} \left(1 + \frac{\Delta \mathcal{A_{\rm LOSA}}}{\mathcal{A}} \right) f^{-7/6} e^{\iota \left[ \Psi (f) + \Delta \Psi_{\rm LOSA} (f) \right]}
\end{equation}
where $\mathcal{A} \propto \sqrt{5/24} \mathcal{M}^{5/6}/D_{\rm L}$ is the amplitude, $\mathcal{M} \equiv M \eta^{3/5}$ is the detector frame chirp mass of the non-accelerating CBC, $D_{\rm L}$ is the luminosity distance, $\Psi (f)$ is the unperturbed phase, $\Delta \mathcal{A_{\rm LOSA}} / \mathcal{A}$ is the relative amplitude correction due to LOSA (see Eq.~A8 of \cite{Vijaykumar_2023}), $\Delta \Psi_{\rm LOSA} (f) \equiv \Delta \Psi_{\rm pp}(f) + \Delta \Psi_{\rm AS} (f) + \Delta \Psi_{\Lambda} (f)$, $\Delta \Psi_{\rm pp} (f)$ is the phase correction due to LOSA containing the higher order non-spinning point-particle terms upto $3.5 {\rm PN}$ and is given by Eq.~A7 of \cite{Vijaykumar_2023} with $\Gamma$ replaced by $\Gamma_1$, while $\Delta \Psi_{\rm AS} (f)$ and $\Delta \Psi_{\Lambda} (f)$ are the same containing the contributions from aligned spin (AS) and tidal effects, respectively. 

Proceeding as in \cite{Vijaykumar_2023}, $\Delta \Psi_{\rm AS} (f)$ and $\Delta \Psi_{\Lambda} (f)$ are derived using the corresponding corrections to the pp expressions of the orbital binding energy and GW flux given in Eq.~3.1 and 3.2 of \cite{PhysRevD.80.084043}. Note, however, that the AS and tidal effects are treated separately.

\subsection{Tidal Effects}\label{subsec:Tidal_eff}
The contributions of the tidal effects to the energy and flux are given by Eq.~6.5b of \cite{PhysRevD.101.064047} and Eq.~6.2 of \cite{PhysRevD.102.044033}, respectively, with $x$ replaced by $v_{\rm u}^2$. Here, $v_{\rm u} \equiv \left(\pi M f_{\rm u}\right)^{1/3}$ and $f_{\rm u}$ is the unperturbed Fourier frequency. Denoting $f_{\rm o}$ as the Doppler-shifted Fourier frequencies, Eq. ~A1-A3 of \cite{Vijaykumar_2023} take the form:
\begin{align}
    \label{eq: dop_shift}
    f_{\rm u} &= f_{\rm o} \left(1 + \Gamma_1 (t_{\rm o} - t_{\rm c}) \right) \\
    dt_{\rm u} &= \frac{dt_{\rm o}}{\left(1 + \Gamma_1 (t_{\rm o} - t_{\rm c}) \right)} \\
    v_{\rm u} &= v_{\rm o} \left(1 + \Gamma_1 (t_{\rm o} - t_{\rm c}) \right)^{1/3}
\end{align}
where $v_{\rm o} = \left(\pi M f_{\rm o}\right)^{1/3}$. Eq.~A4 of \cite{Vijaykumar_2023} becomes:
\begin{equation}
    \label{eq: dvdt}
    \frac{dv_{\rm o}}{dt_{\rm o}} = - \frac{\Gamma_1 v_{\rm o}}{3} + \left(1 + \Gamma_1 (t_{\rm o} - t_{\rm c}) \right)^{-4/3} \frac{dv_{\rm u}}{dt_{\rm u}}
\end{equation}
We use Eq. 3.3b of \cite{PhysRevD.80.084043} to first get $\frac{dv_{\rm u}}{dt_{\rm u}}$ and then follow the procedure mentioned in \cite{Vijaykumar_2023} for further calculations. The expressions of time and GW phase as a function of frequency are given by (Ref. Eq.~A5-A6 of \cite{Vijaykumar_2023}:
\begin{widetext}
    \begin{multline}
        \label{eq: negcoal_tidal}
        (t - t_{\rm c})_{\Lambda} = -\frac{325 \Gamma_1 M^2}{393216 \eta ^2 v^{16}} \Biggl[ \Biggl\{\left(\frac{144}{13 \eta }+\frac{3168}{13}\right)\tilde{\mu}_{+}^{(2)} + \frac{144}{13 \eta } \delta  \tilde{\mu}_{-}^{(2)} \Biggr\} v^{10} + \Biggl\{\left(\frac{4132 \eta }{13}+\frac{5429}{273 \eta } +\frac{149468}{273}\right) \tilde{\mu}_{+}^{(2)} \\ + \left(\frac{5429}{273 \eta }+\frac{1059}{13}\right) \delta \tilde{\mu}_{-}^{(2)} -\left(\frac{8}{39 \eta }-\frac{5536}{13} \right) \tilde{\sigma}_{+}^{(2)} -\frac{8}{39 \eta } \delta  \tilde{\sigma}_{-}^{(2)} \Biggr\} v^{12} + \Biggl\{ - \left( \frac{30440 \eta ^2}{39} + \frac{2842727 \eta }{1638} + \frac{67496677}{825552 \eta } + \frac{43532659}{15288} \right) \tilde{\mu}_{+}^{(2)} \\ - \left(\frac{11264 \eta}{39} + \frac{67496677}{825552 \eta } - \frac{2675147}{3276} \right) \delta \tilde{\mu}_{-}^{(2)} -\frac{3200}{13} \tilde{\mu}_{+}^{(3)} - \left(\frac{48832 \eta}{39} - \frac{4168}{2457 \eta } + \frac{9270256}{2457} \right) \tilde{\sigma}_{+}^{(2)} \\ + \left(\frac{4168}{2457 \eta } - \frac{6872}{13} \right) \delta \tilde{\sigma}_{-}^{(2)} \Biggr\} v^{14} + \Biggl\{ \left(\frac{19424  \eta}{455}+\frac{39028}{91 \eta }+\frac{4977144}{455} \right) \tilde{\mu}_{+}^{(2)} + \left(\frac{39028}{91 \eta }+\frac{535156}{455} \right) \delta \tilde{\mu}_{-}^{(2)} \\ + \left(\frac{10100224}{1365}-\frac{5504}{1365 \eta } \right) \tilde{\sigma}_{+}^{(2)} -\frac{5504 \delta}{1365 \eta } \tilde{\sigma}_{-}^{(2)} \Biggr\} \pi v^{15} \Bigg]
    \end{multline}
    
    \begin{multline}
        \label{eq: phi_tidal}
        \phi_{\Lambda} = -\frac{25 \Gamma_1 M}{24576 \eta ^2 v^{13}} \Biggl[ \left\{ \left( \frac{18}{\eta }+396 \right) \tilde{\mu}_{+}^{(2)} + \frac{18}{\eta } \delta \tilde{\mu}_{-}^{(2)} \right\} v^{10} + \Biggl\{ \left( 1033 \eta + \frac{5429}{84 \eta } + \frac{37367}{21} \right) \tilde{\mu}_{+}^{(2)} + \left( \frac{5429}{84 \eta }+\frac{1059}{4} \right) \delta \tilde{\mu}_{-}^{(2)} \\ + \left(1384 - \frac{2}{3 \eta} \right) \tilde{\sigma}_{+}^{(2)} - \frac{2}{3 \eta } \delta  \tilde{\sigma}_{-}^{(2)} \Biggr\} v^{12} + \Biggl\{ \left(\frac{3805 \eta^2}{3}+\frac{2842727 \eta}{1008}+\frac{67496677}{508032 \eta }+\frac{43532659}{9408} \right) \tilde{\mu}_{+}^{(2)} \\ + \left( \frac{1408 \eta}{3}+\frac{67496677}{508032 \eta }+\frac{2675147}{2016} \right) \delta  \tilde{\mu}_{-}^{(2)} + 400 \tilde{\mu}_{+}^{(3)} + \left( \frac{6104 \eta}{3}-\frac{521 }{189 \eta }+\frac{1158782}{189} \right) \tilde{\sigma}_{+}^{(2)} \\ + \left( 859 - \frac{521 }{189 \eta } \right) \delta  \tilde{\sigma}_{-}^{(2)} \Biggr\} v^{14} + \Biggl\{ - \left(\frac{607}{35} + \frac{9757}{56 \eta } + \frac{622143}{140} \right) \tilde{\mu}_{+}^{(2)} - \left(\frac{133789}{280} + \frac{9757}{56 \eta} \right) \delta  \tilde{\mu}_{-}^{(2)} \\ + \left( \frac{172}{105 \eta } - \frac{315632}{105} \right) \tilde{\sigma}_{+}^{(2)} + \frac{172}{105 \eta } \delta  \tilde{\sigma}_{-}^{(2)} \Biggr\} \pi v^{15}  \Biggr]
    \end{multline}
\end{widetext}
These are only the additional corrections resulting from the tidal effects---the complete expressions for $t - t_{\rm c}$ and $\phi$ can be obtained by adding these terms to the right-hand sides of Eq.~A4 and A5 of \cite{Vijaykumar_2023}, respectively. Finally, the contribution of the tidal effects to the phase correction is given by:
\begin{widetext}
    \begin{multline}
        \label{eq: phase_corr_tidal}
        \Delta \Psi_{\Lambda} (f) = \frac{25 \Gamma_1  M}{65536 \eta^2 v^{13}} \Biggl[ \Biggl\{ \left( \frac{48}{\eta }+1056 \right) \tilde{\mu}_{+}^{(2)} + \frac{48 \delta}{\eta} \tilde{\mu}_{-}^{(2)} \Biggr\} v^{10} +  \Biggl\{ \left( 4132 \eta + \frac{5429}{21 \eta} + \frac{149468}{21} \right) \tilde{\mu}_{+}^{(2)} + \left( \frac{5429}{21 \eta} + 1059 \right) \delta \tilde{\mu}_{-}^{(2)} \\ + \left( 5536 - \frac{8}{3 \eta} \right) \tilde{\sigma}_{+}^{(2)} - \frac{8}{3 \eta} \delta \tilde{\sigma}_{-}^{(2)} \Biggr\} v^{12} + \Biggl\{ \left( \frac{30440 \eta^2}{3} + \frac{2842727 \eta}{126} + \frac{67496677}{63504 \eta } + \frac{43532659}{1176} \right) \tilde{\mu}_{+}^{(2)} \\ + \left( \frac{11264 \eta}{3}+\frac{67496677}{63504 \eta } + \frac{2675147}{252} \right) \delta  \tilde{\mu}_{-}^{(2)} + 3200 \tilde{\mu}_{+}^{(3)} + \left( \frac{48832 \eta }{3}-\frac{4168}{189 \eta }+\frac{9270256}{189} \right) \tilde{\sigma}_{+}^{(2)} \\ + \left( 6872 -\frac{4168}{189 \eta } \right) \delta  \tilde{\sigma}_{-}^{(2)} \Biggr\} v^{14} + \Biggl\{ - \left( \frac{9712\eta}{35}  + \frac{19514}{7 \eta } + \frac{2488572}{35} \right) \tilde{\mu}_{+}^{(2)} - \left( \frac{267578}{35} + \frac{19514}{7 \eta } \right) \delta  \tilde{\mu}_{-}^{(2)} \\ + \left( \frac{2752}{105 \eta } - \frac{5050112 }{105} \right) \tilde{\sigma}_{+}^{(2)} + \frac{2752}{105 \eta } \delta  \tilde{\sigma}_{-}^{(2)} \Biggr\} \pi v^{15} \Biggr]
    \end{multline}
\end{widetext}
where $v \equiv \left(\pi M f\right)^{1/3}$, $\tilde{\mu}_{\pm}^{(l)}$ and $\tilde{\sigma}_{\pm}^{(l)}$ are given by Eq.~6.2 of \cite{PhysRevD.101.064047} and $\delta = \frac{m_1 - m_2}{M}$, $m_1$ and $m_2$ being the detector frame masses of the primary and secondary objects in the binary. Specifically, the contribution of the mass-type quadrupolar second Love number to the phase correction is given by:
\begin{widetext}
    \begin{multline}
        \label{eq: phase_corr_tiadl_l2}
        \Delta \Psi_{\Lambda, 2} (f) = \frac{25 \Gamma_1  M}{65536 \eta^2 v^{13}} \sum_{A = 1, 2} \Lambda_A X_A^4 \biggl[ (576 - 528 X_A) v^{10} + \left( 2066 X_A^3 - 5191 X_A^2 + \frac{4021 X_A}{42} + \frac{46029}{14} \right) v^{12} \\ + \left( -\frac{15220 X_A^5}{3}+\frac{56924 X_A^4}{3}-\frac{3358153 X_A^3}{252}-\frac{576641 X_A^2}{28}+\frac{48108089 X_A}{7056}+\frac{226452487}{15876} \right) v^{14} \\ + \left( -\frac{4856 X_A^3}{35} + \frac{55458 X_A^2}{7}+\frac{838063 X_A}{35}-\frac{172581}{5}  \right) \pi v^{15} \biggr]
    \end{multline}
\end{widetext}
where $X_A = \frac{m_A}{M}$ and $\Lambda_A$ is the (dimensionless) tidal deformability of the object $A$. In order to arrive at Eq.~\ref{eq: phase_corr_tiadl_l2}, one needs to first use Eq.~6.2 and 5.3 of \cite{PhysRevD.101.064047} to arrive at $\tilde{\mu}_{\pm}^{(2)} = \frac{1}{2M^5} \left( \frac{X_2}{X_1} \mu_1^{(2)} \pm \frac{X_1}{X_2} \mu_2^{(2)} \right)$, use Eq. ~2.8 of \cite{PhysRevD.101.064047} together with Eq. ~1.4 of \cite{PhysRevD.102.044033} to obtain $\mu_A^{(2)} = \Lambda_A M^5 X_A^5$, and then finally substitue the relation $\tilde{\mu}_{\pm}^{(2)} = \frac{1}{2} \sum_{A = 1, 2} (1 - X_A) X_A^4 \Lambda_A$ in Eq.~\ref{eq: phase_corr_tidal}. Here, $\mu_1^{(2)}$ and $\mu_2^{(2)}$ are the tidal polarizability coefficients and depend on the mass-type quadrupolar second
Love numbers through Eq.~1.6a of \cite{PhysRevD.102.044033}.

\subsection{Aligned Spin}\label{subsec: AS_eff}
To calculate the contributions of AS effects to the phase correction, we follow the same approach as in \ref{subsec:Tidal_eff}. In this case, the expressions of the energy and flux are given by Eqs~6.18 and 6.19 of \cite{PhysRevD.79.104023, PhysRevD.84.049901} (see also~\cite{Bohe:2015ana}) with $v$ replaced by $v_{\rm u}$. Eq.~\ref{eq: negcoal_tidal} and \ref{eq: phi_tidal} in this case become:
\begin{widetext}
    \begin{multline}
        \label{eq: negcoal_AS}
        (t - t_{\rm c})_{\rm AS} = -\frac{325 \Gamma_1 M^2}{393216 \eta ^2 v^{16}} \Biggl[ \Biggl\{ \left(\frac{904 }{39} - \frac{608 \eta}{39} \right) \chi_{\rm s} + \frac{904}{39} \delta  \chi_{\rm a} \Biggr\} v^3 + \Biggl\{ \left(\frac{9 \eta }{13}-\frac{729}{52}\right) \chi_{\rm s}^2 + \left(\frac{720 \eta }{13}-\frac{729}{52}\right) \chi_{\rm a}^2 \\ - \frac{729}{26} \delta  \chi_{\rm a} \chi_{\rm s} \Biggr\} v^4 + \Biggl\{ \left(-\frac{43072 \eta ^2}{585}-\frac{777856 \eta }{4095}+\frac{1201204}{4095}\right) \chi_{\rm s} + \left(\frac{46496 \eta }{585}+\frac{1201204}{4095}\right) \delta  \chi_{\rm a} \Biggr\} v^5 \\ + \Biggl\{ \left(\frac{424753 \eta ^2}{2925}-\frac{5629937 \eta }{11700}+\frac{295582}{2925}\right) \chi_{\rm s}^2 + \left(\frac{994739 \delta ^2}{5850}+504 \eta ^2+\frac{65503 \eta }{468}-\frac{5381}{78}\right) \chi_{\rm a}^2 -\frac{86716 \pi}{325} \delta  \chi_{\rm a} \\ + \left(\left(\frac{591164}{2925}-\frac{3610481 \eta }{5850}\right) \delta \chi_{\rm a} + \frac{177296 \pi  \eta }{975}-\frac{86716 \pi }{325}\right) \chi_{\rm s}  \Biggr\} v^6 + \Biggl\{ \left(-\frac{1424 \eta ^2}{65}+\frac{21628 \eta }{65}-\frac{25158}{65}\right) \chi_{\rm s}^3 \\ + \left(\frac{19744 \eta }{13} - \frac{25158}{65}\right) \delta \chi_{\rm a}^3  + \left(\left(\frac{41344\eta }{65}-\frac{75474}{65}\right)  \delta  \chi_{\rm a}  -\frac{1152 \pi  \eta}{65} +\frac{17568 \pi }{65}\right) \chi_{\rm s}^2 + \left(\frac{17568}{65}-\frac{13824  \eta }{13}\right) \pi \chi_{\rm a}^2 \\ + \Biggl(\left(-\frac{239444 \delta ^2}{585}-\frac{15488 \eta ^2}{13}+\frac{1919524 \eta }{585}-\frac{439822}{585}\right) \chi_{\rm a}^2 + \frac{35136 \pi}{65} \delta  \chi_{\rm a} - \frac{123842 \eta ^3}{585}-\frac{12880727 \eta ^2}{24570} \\ -\frac{10028546561 \eta }{2063880}+\frac{30198206713}{8255520}\Biggr) \chi_{\rm s} + \left(\frac{194267 \eta ^2}{1170}-\frac{3042577  \eta }{3780}+\frac{30198206713 }{8255520}\right) \delta \chi_{\rm a} \Biggr\} v^7 \Biggr]
    \end{multline}
    \begin{multline}
        \label{eq: phi_AS}
        \phi_{\rm AS} = -\frac{25 \Gamma_1 M}{24576 \eta ^2 v^{13}} \Biggl[ \Bigg\{ \left(\frac{1469}{60}-\frac{247 \eta }{15}\right) \chi_{\rm s} + \frac{1469}{60} \delta \chi_{\rm a} \Biggr\} v^3 + \Biggl\{ \left(\frac{3 \eta }{4}-\frac{243}{16}\right) \chi_{\rm s}^2 + \left(60 \eta -\frac{243}{16}\right) \chi_{\rm a}^2 - \frac{243}{8} \delta  \chi_{\rm a} \chi_{\rm s} \Biggr\} v^4 \\ + \Biggl\{ \left(-\frac{7403 \eta ^2}{90}-\frac{66847 \eta }{315}+\frac{3303311}{10080}\right) \chi_{\rm s} + \left(\frac{15983 \eta }{180}+\frac{3303311}{10080}\right) \delta \chi_{\rm a}  \Biggr\} v^5 + \Biggl\{ \left(\frac{60679 \eta ^2}{360}-\frac{5629937 \eta }{10080}+\frac{21113}{180}\right) \chi_{\rm s}^2 \\ + \left(\frac{994739 \delta ^2}{5040}+585 \eta ^2+\frac{327515 \eta }{2016}-\frac{26905}{336}\right) \chi_{\rm a}^2 + \left(\left(\frac{21113}{90}-\frac{515783 \eta }{720}\right) \delta  \chi_{\rm a} + \frac{3166 \pi  \eta }{15}-\frac{3097 \pi }{10}\right) \chi_{\rm s} - \frac{3097 \pi }{10} \delta  \chi_{\rm a}  \Biggr\} v^6 \\ + \Biggl\{ \left(-\frac{267 \eta ^2}{10}+\frac{16221 \eta }{40}-\frac{37737}{80}\right) \chi_{\rm s}^3 + \left(1851 \delta  \eta -\frac{37737 \delta }{80}\right) \chi_{\rm a}^3 + \left(\left(\frac{3876 \eta }{5}-\frac{113211 }{80}\right) \delta \chi_{\rm a} - \frac{108 \pi  \eta }{5} +\frac{1647 \pi }{5}\right) \chi_{\rm s}^2  \\ + \left(\frac{1647}{5}-1296 \eta \right) \pi \chi_{\rm a}^2 + \Biggl(\left(-\frac{59861 \delta ^2}{120}-1452 \eta ^2+\frac{479881 \eta }{120}-\frac{219911}{240}\right) \chi_{\rm a}^2 + \frac{3294 \pi}{5} \delta  \chi_{\rm a} - \frac{61921 \eta ^3}{240}-\frac{12880727 \eta ^2}{20160} \\ - \frac{10028546561 \eta }{1693440}+\frac{30198206713}{6773760}\Biggr)\chi_{\rm s} + \left(\frac{194267 \eta^2}{960}-\frac{39553501 \eta }{40320}+\frac{30198206713  }{6773760}\right) \delta \chi_{\rm a} \Biggr\} v^7  \Biggr]
    \end{multline}
\end{widetext}
As with the corresponding expressions involving tidal effects, these too are only the additional corrections resulting from AS effects, and the complete expressions for $t - t_{\rm c}$ and $\phi$ can be obtained by adding these terms to the right-hand sides of Eq.~A4 and A5 of \cite{Vijaykumar_2023}, respectively. The contribution of AS to the phase correction is given by:
\begin{widetext}
    \begin{multline}
        \label{eq: phase_corr_AS}
        \Delta \Psi_{\rm AS} (f) = \frac{25 \Gamma_1  M}{65536 \eta^2 v^{13}} \Biggl[ \Biggl\{ \left(\frac{452}{15}-\frac{304 \eta }{15} \right) \chi_{\rm s} + \frac{452}{15}\delta  \chi_{\rm a} \Biggr\} v^3 + \Biggl\{ \left(\eta -\frac{81}{4}\right) \chi_{\rm s}^2 + \left(80 \eta -\frac{81}{4}\right) \chi_{\rm a}^2 - \frac{81}{2} \delta  \chi_{\rm a} \chi_{\rm s} \Biggr\} v^4 \\ + \Biggl\{ \left(-\frac{5384 \eta ^2}{45}-\frac{97232 \eta }{315}+\frac{300301}{630})\right) \chi_{\rm s} + \left(\frac{5812 \eta }{45}+\frac{300301}{630}\right) \delta \chi_{\rm a} \Biggr\} v^5 + \Biggl\{ \left(\frac{60679 \eta ^2}{225}-\frac{5629937 \eta }{6300}+\frac{42226}{225}\right) \chi_{\rm s}^2 \\ + \left(936 \eta ^2-\frac{2106779 \eta }{2100}+\frac{42226}{225}\right) \chi_{\rm a}^2 + \left(\left(\frac{84452}{225} - \frac{515783 \eta }{450}\right) \delta \chi_{\rm a} + \frac{25328 \pi  \eta }{75} - \frac{12388 \pi }{25}\right) \chi_{\rm s} -\frac{12388 \pi}{25} \delta  \chi_{\rm a} \Biggr\} v^6 \\ + \Biggl\{ \left(-\frac{712 \eta ^2}{15}+\frac{10814 \eta }{15}-\frac{4193}{5}\right) \chi_{\rm s}^3 + \left(\frac{9872 \delta  \eta }{3}-\frac{4193 \delta }{5}\right) \chi_{\rm a}^3 +  \left(\left(\frac{20672 \eta }{15} - \frac{12579}{5}\right) \delta \chi_{\rm a} -\frac{1}{5} 192 \pi  \eta +\frac{2928 \pi }{5}\right) \chi_{\rm s}^2 \\ + \left(\frac{2928 \pi }{5} - 2304 \pi  \eta \right) \chi_{\rm a}^2 + \Biggl(\frac{5856 \pi}{5} \delta \chi_{\rm a} - \frac{61921 \eta ^3}{135}+\left(-\frac{7744 \eta ^2}{3}+\frac{31970 \eta }{3}-\frac{12579}{5}\right) \chi_{\rm a}^2-\frac{12880727 \eta ^2}{11340} \\ -\frac{10028546561 \eta }{952560}+\frac{30198206713}{3810240}\Biggr) \chi_{\rm s} + \left(\frac{194267 \eta^2}{540}-\frac{39553501 \eta }{22680}+\frac{30198206713}{3810240}\right) \delta \chi_{\rm a} \Biggr\} v^7  \Biggr]
    \end{multline}
\end{widetext}
where $\chi_{\rm s} = \frac{1}{2}(\chi_1 + \chi_2)$, $\chi_{\rm a} = \frac{1}{2}(\chi_1 - \chi_2)$, and $\chi_1$ and $\chi_2$ are the dimensionless spins of primary and secondary objects projected onto the angular momentum of the binary, respectively. Note that this is consistent with \cite{PhysRevD.110.083008}, which also computes the aligned spin corrections for accelerating CBCs and presents a scenario of detecting LOSA with next generation ground-based detectors. The Mathematica notebooks used for derivations can be found in Ref.~\cite{SM_MP2}.

The phase corrections in \cite{Vijaykumar_2023} and this paper have been derived assuming the \textsc{TaylorF2} model, which is inspiral only, but we will be adding these corrections to the IMR waveforms that have merger as well as ringdown. However, we do not expect this to bias the observed parameters simply because the leading order phase correction appears at $-4{\rm PN}$, which will become highly suppressed at the higher frequencies. Additionally, the merger and rigndown parts of the signal will become insensitive to the LOSA effects because the signal duration $\Delta t$ after the last stable orbit becomes too small. Hence, $\Gamma_1 \Delta t$ will become almost constant, leading again to a constant redshift.

\section{Region of validity of the approximation}\label{sec:approx_validity}
To ascertain where the condition $\vert \Gamma_1 (t_{\rm o} - t_{\rm c}) \vert \ll 1$ used in Eq.~\eqref{eq: redshifted_mass} is met, we perform estimate the Fisher Matrix over a grid of detector frame component masses on the parameter space $\{\ln D_{\rm L},\, \ln \mathcal{M},\, \ln \eta,\, \Gamma_1 \}$. We use the aLIGO PSD\footnote{Available at \href{https://dcc.ligo.org/LIGO-T2000012-v2/public}{https://dcc.ligo.org/LIGO-T2000012-v2/public} with the name \textit{aligo\_O4high.txt}.} and set the lower and upper cut-off frequencies to $10.22{\rm Hz}$ and innermost stable circular orbit frequency $f_{\rm ISCO}$, respectively. We then rescale the constraints on $\Gamma_1$ to optimal signal-to-noise-ratio (SNR) 10 and multiply it by the signal duration $t_{\rm sd}$, obtained by subtracting the times until coalescence evaluated at the lower and upper cut-off frequencies, to get the quantity $ \Delta \Gamma_1 t_{\rm sd}$.

\begin{figure*}[!htbp]
    \centering
    \includegraphics[width=0.475\linewidth]{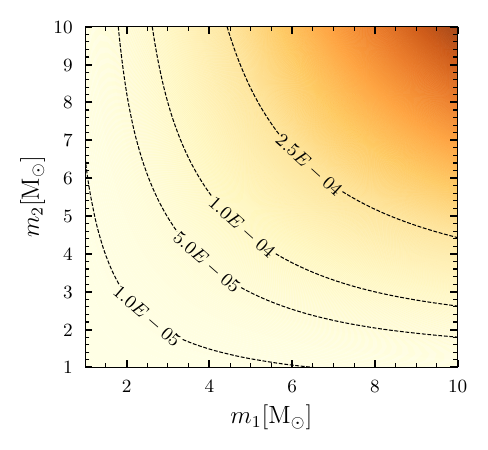}
    \includegraphics[width=0.475\linewidth]{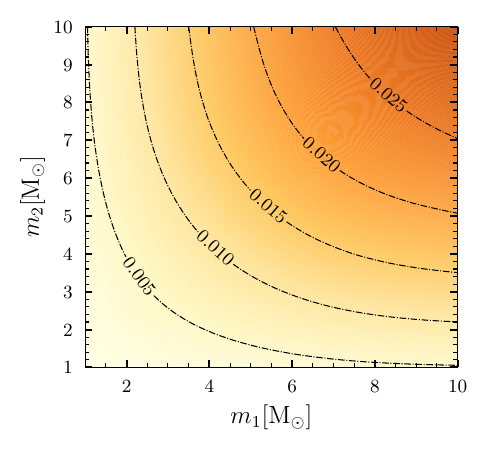}
    \caption{\textbf{Left Panel}: The precision in the measurement of $\Gamma_1$ at SNR 10 by aLIGO over a grid of detector frame component masses. The dashed lines represent the contours of constant $\Delta (\Gamma_1)$. \textbf{Right Panel}: The validity condition of the approximation Eq. \ref{eq: redshifted_mass} at SNR 10 for aLIGO over a grid of detector frame component masses. The dash-dotted lines represent the contours of constant $ \Delta \Gamma_1 t_{\rm sd}$.} 
    \label{fig:dabyc_cond}
\end{figure*}

Figure \ref{fig:dabyc_cond} shows the constraints on $\Gamma_1$ (\textit{left panel}) and the quantity $\vert \Delta \Gamma_1 t_{\rm sd} \vert$ (\textit{right panel}) for aLIGO at optimal SNR 10 over a grid of detector frame component masses. The best constraints on $\Gamma_1$ are of $\mathcal{O}(10^{-6} \,{\rm s}^{-1})$ while the worst constraints on the same are of $\mathcal{O}(10^{-3}\,{\rm s}^{-1})$. In the same parameter space quantity $\vert \Delta (\Gamma_1) t_{\rm sd} \vert$ varies over a range $\mathcal{O}(10^{-3}) - \mathcal{O}(10^{-1})$. 

Assuming that we set a threshold of $0.01$\footnote{While somewhat arbitrary, this number has been chosen to ensure the validity of Newtonian kinematics, and that special relativistic effects can be ignored.} on $\vert \Gamma_1 (t_{\rm o} - t_{\rm c}) \vert$ (to call it $\ll 1$), 
Figure \ref{fig:dabyc_cond} motivates us to summarily reject all events with total detector frame mass (median value) $> 10 \, {\rm M}_{\odot}$ as unsuitable for our analysis. Those that satisfy the total mass cut are analyzed to search for signatures of LOSA. However, if the upper limit on $\Gamma_1$ at $90\%$ is found to violate $\vert \Gamma_1 (t_{\rm o} - t_{\rm c}) \vert$ < 0.01, with $(t_{\rm o} - t_{\rm c})$ estimated using the Newtonian order approximation and median value for the chirp mass, the constraints estimated are deemed unreliable. Additional refinements on the selection criteria are presented in later sections, and the entire list of criteria is summarized in Section~\ref{sec:sel-crit}.

\section{LOSA parameter inference}\label{sec:LOSA_PE}
Before proceeding to waveform systematics, we first perform zero-noise injections for a non-accelerating and an accelerating GW170817-like non-spinning BNS with \textsc{IMRPhenomXP\_NRTidalv2} \cite{Colleoni:2023czp} as the base waveform. We assume the LVK network operating at O4 sensitivity\footnote{We use the same detector configurations for all the studies in this paper and use the PSDs available at \href{https://dcc.ligo.org/LIGO-T2000012-v2/public}{https://dcc.ligo.org/LIGO-T2000012-v2/public}.} to ensure that the pipeline is working as intended. For the non-accelerating injection, we set the injection parameters to: $\{ \mathcal{M} = 1.197 \, {\rm M}_{\odot},\, \eta = 0.844,\, D_{\rm L} = 37 \, {\rm Mpc},\, \Lambda_1 = 250,\, \Lambda_2 = 400,\, {\rm RA} = 3.401,\, {\rm DEC} = -0.335,\, \theta_{\rm JN} = 1.048,\,  \psi = 0.054,\, \phi = 6.052,\, t_{\rm c} = {\rm GPS \, time} \, (1187008882.43),\, \Gamma_1 = 0 \}$, while for the accelerating injection we assume the binary's centre of mass to be accelerating with $a = 3\,{\rm km/s^2}$ or $\Gamma_1 = 10^{-5} \, {\rm s}^{-1}$. We set the signal duration to 512 seconds for all runs carried out in this paper, unless specified otherwise. We estimate source properties using the  \textsc{Dynesty}~\cite{2020MNRAS.493.3132S} sampler implemented in \textsc{Bilby}~\cite{bilby_paper}. To speed up our likelihood evaluations, we use the distance~\cite{Singer:2015ema} and phase marginalized~\cite{Veitch:2014wba} \textsc{Relative Binning Likelihood}~\cite{Cornish:2021lje, Zackay:2018qdy, Krishna:2023bug} to recover the injected parameters using the priors listed in Table \ref{tab:priors} --- unless otherwise mentioned, this method is used for all analyses in this work. However, for simplicity, we fix the geocentric time $t_{\rm c}$ during recovery, and keep it the same for all the studies in this paper. Furthermore, we also keep the right ascension ${\rm RA},$ declination ${\rm DEC},$ the inclination angle (between the total angular momentum and the line of sight) $\theta_{\rm JN},$ polarization angle $\psi,$ and phase at the reference frequency $\varphi$ (set to $20 {\rm Hz}$ across all injection studies in this paper) the same in all studies in this paper except in subsection \ref{sec:GW190814_like_analysis}. 

We repeat this process for an aligned-spinning binary with GW170817-like masses using \textsc{IMRPhenomXAS}\footnote{Note that throughout this paper, we denote \textsc{IMRPhenomXP}~\cite{Pratten:2020ceb} and \textsc{IMRPhenomXPHM}~\cite{Pratten:2020ceb}, with aligned spin settings, by \textsc{IMRPhenomXAS}~\cite{Pratten:2020fqn} and \textsc{IMRPhenomXHM}~\cite{Garcia-Quiros:2020qpx}, respectively.} as the base waveform where we set: $\{\chi_1 = 0.01,\, \chi_2 = 0.009,\, D_{\rm L} = 80 \, {\rm Mpc} \}$ with the rest of the parameters remaining the same as the BNS injections except that $\Lambda_1$ and $\Lambda_2$ are set to $0$ this time.

\begin{figure}
    \centering
    \includegraphics[width=1\linewidth]{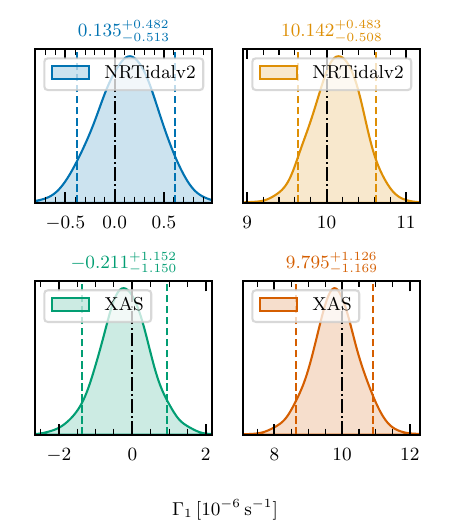}
    \caption{\textbf{Top Panel}: The $1d$ posteriors of $\Gamma_1$ for GW170817 like non-spinning BNS injections using \textsc{IMRPhenomXP\_NRTidalv2} as the base waveform. \textbf{Bottom Panel}: The same for aligned-spin binary injections using \textsc{IMRPhenomXAS} as the base waveform. The \textit{left panels} show the recovery of LOSA for a non-accelerating system. In contrast, the \textit{right panels} show the recovery of the same for a system accelerating with $3 \, {\rm km/s^{2}}$ i.e., $a/c$ being $10^{-5} \, {\rm s}^{-1}$. The dashed lines represent the $90\%$ confidence intervals (CI), while the black dash-dotted lines represent the injected values. The values in the title correspond to the median and $90\%$ CI values. Note that we follow this convention of representing the CIs and injected values throughout this paper.}
    \label{fig:GW170817_like_inj}
\end{figure}

Figure \ref{fig:GW170817_like_inj} shows the recovered $1d$ posteriors of LOSA for these four injections and Table~\ref{tab: snr_bns_inj} shows the corresponding matched-filter network SNRs --- henceforth, referred as SNRs for simplicity. The true value of $\Gamma_1$ is recovered at $90\%$ confidence in all cases. These results act as an important sanity check for \textsc{LOSA-pipe} and suggest that, when obvious sources of systematics are missing, \textsc{LOSA-pipe} recovers LOSA as expected. 

\begin{table}[h]
    \centering
    \begin{tabular}{|c|c|}
        \hline 
        \textbf{Parameter} & \textbf{Prior} \\
        \hline \hline
        LOSA $\Gamma_1 [{\rm s}^{-1}]$   & $\mathcal{U}$(-0.1, 0.1)   \\
        \hline
        Chirp Mass $\mathcal{M} [\rm M_{\odot}]$   & $\mathcal{U}$(1.13715, 1.25685) \\
        \hline
        Mass ratio $q$   & $\mathcal{U}$(0.125, 1)  \\
        \hline
        Luminosity Distance $D_{\rm L} [\, {\rm Mpc}]$  & PL(10, 500) $\propto D_{\rm L}^2$ \\
        \hline 
        $\rm RA$ & $\mathcal{U}$(0, $2\pi$) \\
        \hline
        $\rm DEC$ & Cosine($-\frac{\pi}{2}$, $\frac{\pi}{2})$ \\
        \hline
        $\theta_{\rm JN}$ & Sine(0, $\pi$) \\
        \hline
        Polarization angle $\psi$ & $\mathcal{U}$(0, $\pi$) \\
        \hline
        Phase at the reference frequency $\phi$ & $\mathcal{U}$(0, $2\pi$) \\
        \hline
        Tidal deformabilities $\Lambda_{1,2}$ & $\mathcal{U}(0, 2000)$ \\
        \hline
        Aligned-spins $\chi_{1,2}$ & $\mathcal{U}(0, 0.05)$ \\
        \hline
    \end{tabular}
    \caption{The table of priors used during PE of all GW170817-like injections: $\mathcal{U}$ refers to uniform priors while PL refers to power law. We fix $\chi_{1,2}$ to zero for BNS injections, $\Lambda_{1,2}$ to zero for aligned-spinning binary injections. Furthermore, we use the same priors for LOSA, RA, DEC, $\theta_{\rm JN}$, $\psi$, and $\phi$ throughout this analysis and will mention the priors on chirp mass, mass ratio, or any other parameter whenever we use a different prior than the ones mentioned in this table.}
    \label{tab:priors}
\end{table}

\begin{table}[h]
    \centering
    \begin{tabular}{|c|c|c|}
    \hline
         \textbf{Base Waveform} & $\Gamma_1 \, [\rm s^{-1}]$ & $\rho^N_{\rm mf}$ \\
         \hline
         \hline
         \textsc{IMRPhenomXP\_NRTidalv2} & 0 & 52.04 \\
         \hline
         \textsc{IMRPhenomXP\_NRTidalv2} & $10^{-5}$ & 52.05 \\
         \hline
         \textsc{IMRPhenomXAS} & 0 & 23.99 \\
         \hline
         \textsc{IMRPhenomXAS} & $10^{-5}$ & 23.98 \\
         \hline
    \end{tabular}
    \caption{SNRs of GW170817-like binary injections.}
    \label{tab: snr_bns_inj}
\end{table}

\section{Potential sources of systematics}\label{sec:pot_sos}
In this section, we explore how systematics associated with missing physics in the LOSA waveform could lead to spurious non-zero LOSA recoveries. We consider several of these ``LOSA mimickers'', starting with precession in subsection \ref{subsec:Prec_as_LM} followed by eccentricity in subsection \ref{subsec:Ecc_as_LM}. These mimickers arise because the phase corrections in this paper 
are derived assuming that only the $(l, m) = (2, 2)$ mode is present, both objects in the binary have aligned spins, and the binary orbits are quasi-circular. We then discuss, in subsection \ref{subsec:TidE_as_LM}, how exclusion of tidal effects can lead to a biased LOSA. Finally, in subsections \ref{subsec:DR_as_LM} and \ref{subsec:MG_as_LM}, we discuss how potential (unconfirmed) beyond-GR effects: dipole radiation and massive graviton, could mimic LOSA. Throughout the analyses in this section, we keep the SNRs high so that the mimicker-related biases are not washed away by large statistical errors.

\subsection{Precession as LOSA mimicker}\label{subsec:Prec_as_LM}
To understand the waveform systematics in LOSA inference arising due to the absence of phase correction for precessing spins, we inject a precessing, non-accelerating, and roughly symmetric\footnote{We choose the mass ratio to be high to make sure that higher modes are not excited.} binary of total mass $9.5 \, {\rm M}_{\odot}$ and mass ratio $0.9$ with $a_1 = 0.5$, $a_2 = 0.4$, $\theta_1 = 1.0472$, $\theta_2 = 0.5236$, $\phi_{12} = 0.2618$, $\phi_{jl} = 0.2618$, (i.e. $\chi_p = 0.43$) and $D_{\rm L} = 50 \, {\rm Mpc}$ using \textsc{SEOBNRv5PHM}~\cite{Ramos-Buades:2023ehm} in zero noise, assuming a signal duration of 128 seconds. Then we perform the PE to recover the parameters using \textsc{IMRPhenomXP} and \textsc{IMRPhenomXAS} as the base waveform using priors mentioned in Table~\ref{tab: snr_prior_prec}.

\begin{table}[h]
    \centering
    \begin{tabular}{|c|c|c|}
    \hline
         \textbf{Base Waveform} & \textbf{Prior} & $\rho^N_{\rm mf}$  \\
         \hline
         \hline
         \multirow{5}{*}{\textsc{IMRPhenomXP}} & $\mathcal{M}:\,\mathcal{U}(3.8, 4.4)$ & \multirow{5}{*}{105.51} \\
                       & $q:\,\mathcal{U}(0.05, 1)$ &  \\
                       & $a_{1,2}:\,\mathcal{U}(0, 0.99)$ & \\
                       & $\theta_{1,2}:\,{\rm Sine}(0, \pi)$ & \\
                       & $\phi_{12,\rm JL}:\,\mathcal{U}(0, 2\pi)$ & \\
        \hline
        \multirow{3}{*}{\textsc{IMRPhenomXAS}} & $\mathcal{M}:\,\mathcal{U}(3.8, 4.4)$ & \multirow{3}{*}{104.35} \\
                       & $q:\,\mathcal{U}(0.05, 1)$ &  \\
                       & $\chi_{1,2}:\,\mathcal{U}(0, 0.9)$ & \\
        \hline
    \end{tabular}
    \caption{Priors used in analysing non-accelerating precessing binary injections and corresponding SNRs.}
    \label{tab: snr_prior_prec}
\end{table}

\begin{figure}[h]
    \centering
    \includegraphics[width=1\linewidth]{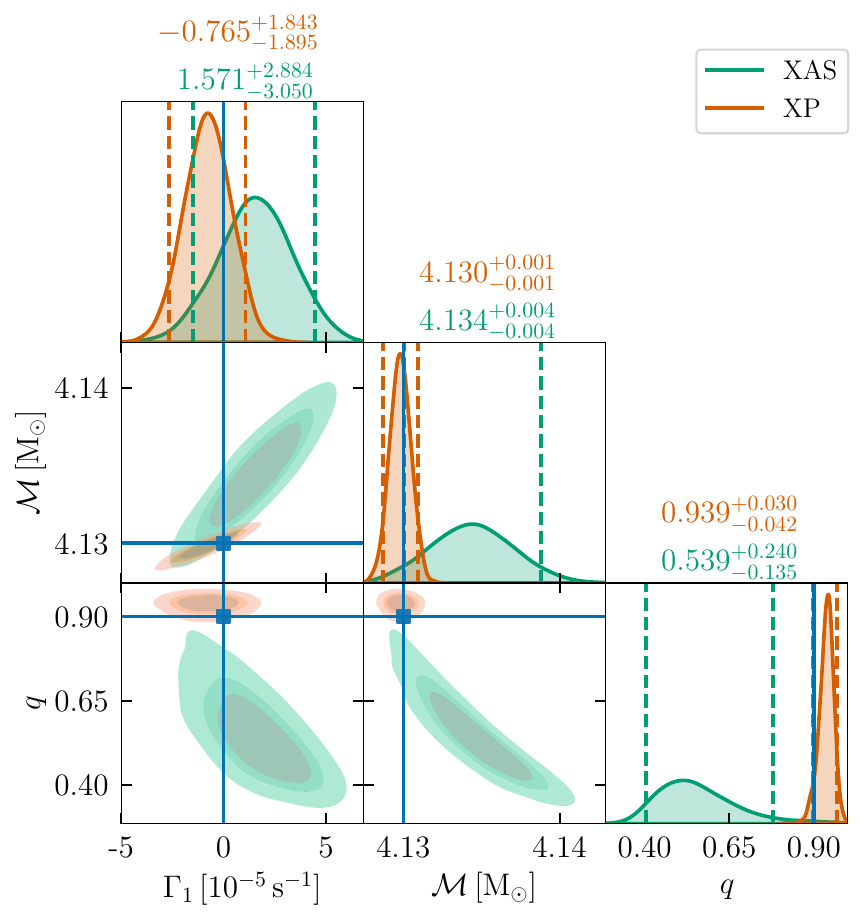}
    \caption{The corner plots of $\Gamma_1$, $\mathcal{M}$, and $q$ highlighting the waveform systematics when using \textsc{IMRPhenomXP} (brown) and \textsc{IMRPhenomXAS} (green) as the base waveforms during recovery of parameters of a non-accelerating binary injected with \textsc{SEOBNRv5PHM}.} 
    \label{fig:corner_plot_comp_prec}
\end{figure}

Figure \ref{fig:corner_plot_comp_prec} shows the corner plots of $\Gamma_1$, $\mathcal{M}$, and $q$ for both recoveries and Table~\ref{tab: snr_prior_prec} shows the corresponding SNRs. Although $\Gamma_1$ is recovered near the $90 \%$ CI, $\mathcal{M}$ and $q$ are not recovered when using the aligned spins\footnote{Note that we will still be fine in the low SNR limit as shown in Figure~\ref{fig:app_low_snr_prec_inj} in Appendix~\ref{app_subsec:low_snr_prec_inj}.}. Furthermore, even though $\Gamma_1$ and $\mathcal{M}$ are recovered within the $90 \%$ CI when using precessing spins, $q$ is still not very well recovered as the injected value is very close to the $90\%$ CI boundary. Therefore, imposing LOSA corrections on \textsc{IMRPhenomXP} for a highly precessing system can bias other parameters, especially in the high SNR limit. To rule out the possibility of systematics between the base waveforms \textsc{SEOBNRv5PHM} and \textsc{IMRPhenomXP} playing a role at such a high SNR, we have checked that \textsc{IMRPhenomXP} can recover all parameters in the absence of LOSA. This is why we propose to add the phase corrections in this paper to a precessing waveform only when the observed GW event has $\chi_{\rm p}\leq 0.4$.

\subsection{Eccentricity as LOSA mimicker}
\label{subsec:Ecc_as_LM}
Orbital eccentricities introduce dephasing in the GW signal that may also be degenerate with the dephasing due to a LOSA. In fact, a quasi-Keplerian computation of the GW phase valid for low eccentricities ($\lesssim0.1$) in a CBC shows that the leading order corrections appear below $-3 \rm PN$ \cite{Moore:2016qxz}. The number of extra cycles contributed by LOSA and by eccentricity to a quasi-circular, non-accelerated binary can be calculated by
\begin{align*}
    \Delta N_{\rm{cyc}}^{\rm{LOSA}} &=\frac{1}{2\pi}\big[\Delta\Psi_{\rm{LOSA}}(f_{\rm{end}})-\Delta\Psi_{\rm{LOSA}}(f_{\rm{start}})\big] \\
    \Delta N_{\rm{cyc}}^{\rm{ecc}} &=\frac{1}{2\pi}\big[\Delta\Psi_{\rm{ecc}}(f_{\rm{end}})-\Delta\Psi_{\rm{ecc}}(f_{\rm{start}})\big] 
\end{align*}
where $f_{\rm{start,end}}$ denote the lower and upper bounds on the GW frequencies, respectively. The 3.5 PN phase differences are calculated as $\Delta\Psi_{\rm{LOSA,ecc}}(f) = \Psi_{\rm{LOSA,ecc}}(f) - \Psi_{\rm{point \, particle}}(f)$. We show the LOSA parameter $\Gamma_1$, and the eccentricities $e_0$ (at reference frequency\footnote{Note that the numbers of extra phase cycles are sensitive to the reference frequency.} of 20 Hz) at which the additional cycles contributed by each are comparable in Table~\ref{tab:phase_difference}. 
\begin{table}[h]
    \centering
    \begin{tabular}{|c|c|c|c|}
    \hline
       $\Gamma_1$ & $\sim10^{-7}$ &  $\sim 10^{-5}$ & $\sim10^{-3}$ \\
       \hline
       $e_0$ & $\sim 0.001$ &  $\sim 0.01$ & $\sim 0.1$ \\
    \hline
    \end{tabular}
    \caption{Orbital eccentricities that can add a similar number of additional cycles to the CBC signal as an accelerated quasi-circular binary.}
    \label{tab:phase_difference}
\end{table}
Note that the quasi-Keplerian formalism remains a valid approximation to compute the phase contribution well into the lower cutoff frequency of the LIGO sensitivity band for CBC masses $\lesssim 10\, \rm M_\odot$ (cf. section III D in \cite{Moore:2016qxz}).  

While the total number of additional cycles suggests interference between the two effects, matched filtering is primarily sensitive to phase modulations in the waveform rather than the accumulated cycle count. Therefore, we further investigate the impact of neglecting eccentricity on LOSA inference using four zero-noise injections of a GW170817-like, non-spinning, non-accelerating, and eccentric binary using the quasi-Keplerian waveform \textsc{TaylorF2Ecc} \cite{TaylorF2Ecc:2016} with $e = \{ 0.005, 0.01, 0.05, 0.1\}$ defined at $20 {\rm Hz}$. We then recover LOSA together with other parameters using \textsc{IMRPhenomXAS} as the base waveform with $\chi_{1,2}$ fixed to zero. 

\begin{table}[h]
    \centering
    \begin{tabular}{|c|c|}
    \hline
         $e$ & $\rho^N_{\rm mf}$ \\
         \hline
         \hline
         0.005 & 44.31 \\
         \hline
         0.01 & 44.29 \\
         \hline
         0.05 & 43.84 \\
         \hline
         0.1 & 41.69 \\
         \hline
    \end{tabular}
    \caption{SNRs of GW170817-like non-spinning eccentric binary injections interpreted as accelerating binaries.}
    \label{tab: snr_ecc}
\end{table}

\begin{figure}
    \centering
    \includegraphics[width=1\linewidth]{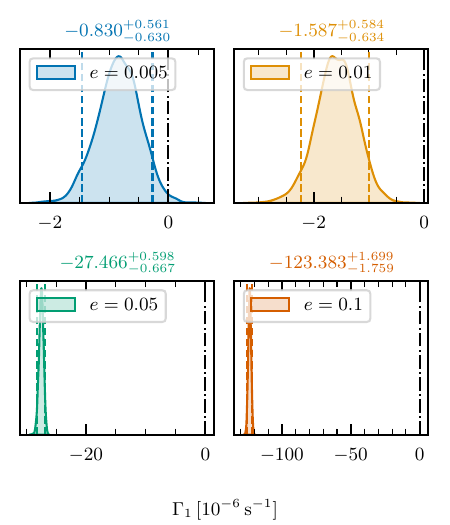}
    \caption{The $1d$ posteriors of $\Gamma_1$ for GW170817 like non-spinning eccentric binary injections recovered with \textsc{IMRPhenomXAS} as the base waveform. \textbf{Top left panel}: $1d$ posterior of LOSA for an eccentricity $e = 0.005$. \textbf{Top right panel}: $1d$ posterior of LOSA for $e = 0.01$. \textbf{Bottom left panel}:$1d$  posterior of LOSA for $e = 0.05$. \textbf{Bottom right panel}: $1d$ posterior of LOSA for $e = 0.1$.} 
    \label{fig:abyc_ecc}
\end{figure}

Figure \ref{fig:abyc_ecc} shows the marginal posteriors of $\Gamma_1$ for the mentioned eccentric injections and Table~\ref{tab: snr_ecc} shows the corresponding SNRs. In all of the cases, $\Gamma_1$ is recovered away from zero, and the bias becomes more severe for high eccentricities. Therefore, an eccentric merger can be confused with an accelerating one and must be followed up with model selection to claim either LOSA or eccentric merger.

\subsection{Tidal Effects as LOSA mimickers}\label{subsec:TidE_as_LM}
To understand the waveform systematics in LOSA inference when we neglect the tidal effects, we inject a non-accelerating GW170817-like BNS considered in section \ref{sec:LOSA_PE} using \textsc{IMRPhenomXP\_NRTIdalv2} in zero noise and then recover $\Gamma_1$ and other parameters using \textsc{IMRPhenomXAS} and \textsc{IMRPhenomXP\_NRTidalv2} as the base waveforms.

\begin{figure}
    \centering
    \includegraphics[width=1\linewidth]{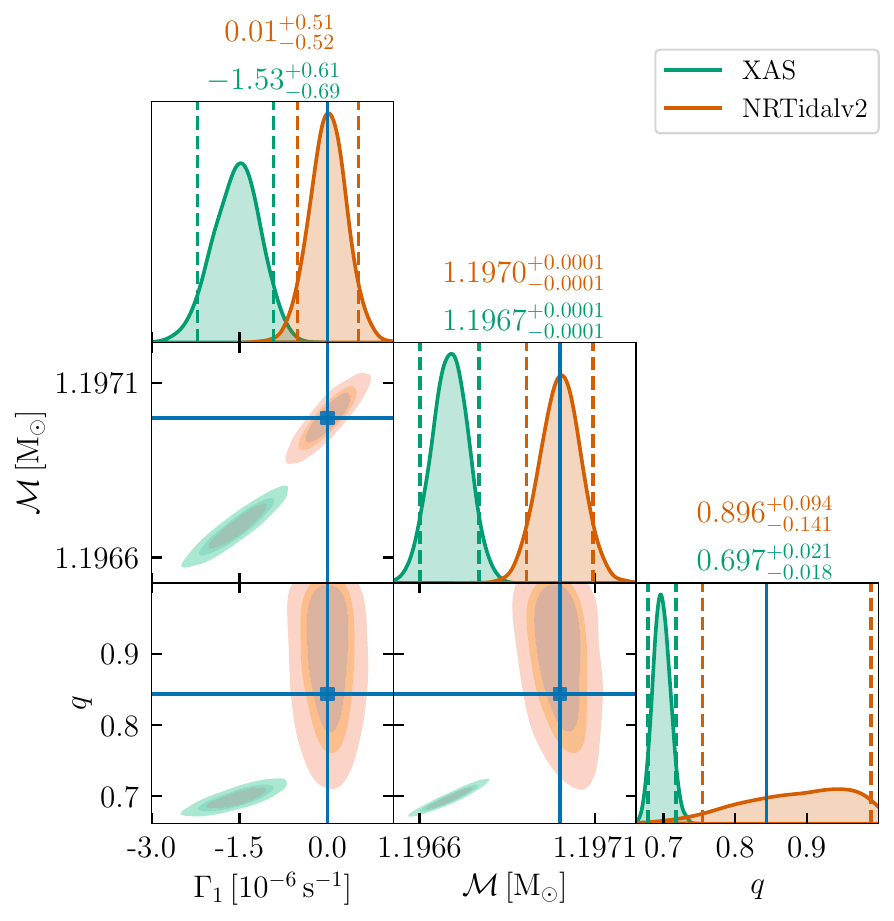}
    \caption{The corner plot of $\Gamma_1$, $\mathcal{M}$, and $q$ highlighting the waveform systematics when using \textsc{IMRPhenomXAS} (green) and \textsc{IMRPhenomXP\_NRTidalv2} (brown) as the base waveforms during recovery of parameters of a non-accelerating GW170817-like signal injected with \textsc{IMRPhenomXP\_NRTidalv2} in zero-noise.}
    \label{fig:corner_plot_comp_tidal}
\end{figure}

\begin{table}[h]
    \centering
    \begin{tabular}{|c|c|}
    \hline
         Base Waveform & $\rho^N_{\rm mf}$ \\
         \hline
         \hline
         \textsc{IMRPhenomXP\_NRTidalv2} & 52.04 \\
         \hline
         \textsc{IMRPhenomXAS} & 50.69 \\
         \hline
    \end{tabular}
    \caption{SNRs of GW170817-like non-accelerating BNS injections interpreted as accelerating binaries.}
    \label{tab: snr_tidal}
\end{table}

Figure \ref{fig:corner_plot_comp_tidal} shows the corner plots of $\Gamma_1$, $\mathcal{M}$, and $q$ for both of these cases and Table~\ref{tab: snr_tidal} shows the corresponding SNRs. We find that none of $\Gamma_1$, $\mathcal{M}$, or $q$ are recovered when using \textsc{IMRPhenomXAS} as the base waveform, while all of these are very well recovered when using \textsc{IMRPhenomXP\_NRTIdalv2}. Therefore, the exclusion of tidal effects in binaries can mimic LOSA.

\subsection{Dipole Radiation as a LOSA mimicker}\label{subsec:DR_as_LM}
Due to the conservation of the matter stress-energy tensor in GR, only quadrupole or higher-order multipole moments contribute to the GW emission. However, there are several beyond-GR theories where the monopole and dipole emissions are also allowed due to the matter stress-energy tensor not being conserved \cite{PhysRevLett.116.241104}. Assuming that $B$ is a theory-dependent parameter that can regulate the strength of the dipole term, the phase correction due to the dipole radiation can be written as \cite{PhysRevLett.116.241104, PhysRevD.86.022004}: 
\begin{equation}
    \label{eq:phase_corr_dr}
    \Delta \Psi_{-1,\rm DR} = - \frac{3}{224 \eta v^7} B  
\end{equation}
where $v$ is the post-Newtonian parameter defined in section \ref{sec:phase_corr_der} and $\eta$ is the symmetric mass ratio.

We add Eq. \ref{eq:phase_corr_dr} to the phase of \textsc{IMRPhenomXAS} and inject a non-accelerating GW170817-like binary in zero-noise by setting \footnote{We choose this value of $B$ to be consistent with the most stringent constraint on it, i.e., $\vert B \vert < 1.9 \times 10^{-3}$ \cite{PhysRevLett.116.241104}.} $B = 10^{-3}$ and then perform the PE to infer LOSA and other parameters using \textsc{IMRPhenomXAS} as the base waveform.

\begin{figure*}
    \centering
    \includegraphics[width=1\linewidth]{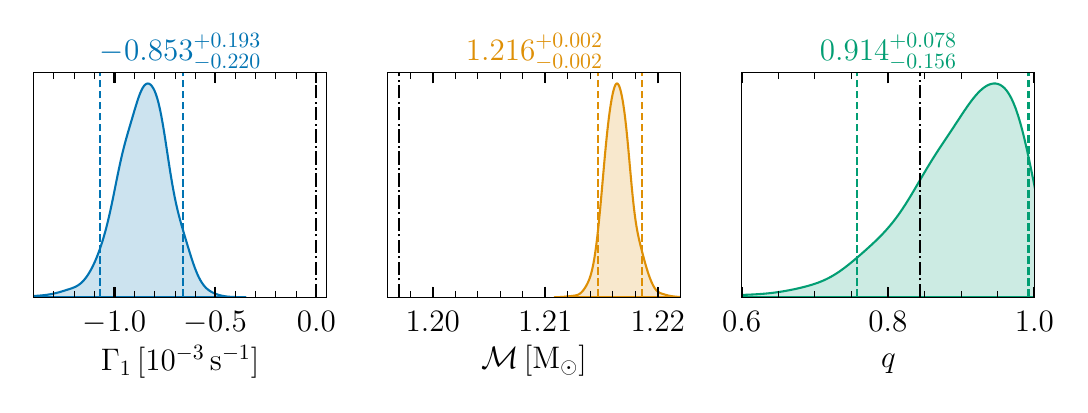}
    \caption{$1d$ posteriors of LOSA (\textit{left panel}), $\mathcal{M}$ (\textit{central panel}), and $q$ (\textit{right panel}) with \textsc{IMRPhenomXAS} as the base waveforms for a non-spinning and non-accelerating GW170817-like CBC at $100 \, {\rm Mpc}$, injected with \textsc{IMRPhenomXAS} modified with contributions from dipole GW emission.}
    \label{fig:post_dr}
\end{figure*}

We find the SNR $\rho_{\rm mf}^N$ to be $10.84$. Figure \ref{fig:post_dr} shows the posteriors of $\Gamma_1$, $\mathcal{M}$, and $q$. Although $q$ is recovered, $\Gamma_1$ and $\mathcal{M}$ both are completely biased. Hence, dipole radiation can mimic LOSA. 

\subsection{Massive Graviton as a LOSA mimicker}\label{subsec:MG_as_LM}
A finite mass of graviton modifies the dispersion relation $E^2 = p^2c^2$ as $E^2 = p^2c^2 + A_0$ where $A_0 = m_{\rm g}^2 c^4$, $m_{\rm g}$ is the mass of the graviton, $E$ is the energy of GWs, and $p$ is the momentum of GWs. It introduces a $1{\rm PN}$ correction to the GW phase, which is given by \cite{PhysRevD.100.104036, PhysRevD.57.2061} 
\begin{equation}
    \label{eq:phase_corr_mg}
    \Delta \Psi_{1, \rm MG} = - {\rm sign}(A_0) \frac{\pi D_{\rm L}^2 (1 + z)}{D_0} \frac{c}{f} \frac{1}{\lambda_0^2}
\end{equation}
where $D_{\rm L}$ is the luminosity distance, $\lambda_0 = hc/\sqrt{\vert A_0 \vert} = h/(m_{\rm g}c)$ is the Yukawa screening length, $h$ is the Planck constant, $z$ is the cosmological redshift of the binary, and $D_0$ is given by Eq. 5 of \cite{PhysRevD.100.104036}.

We add Eq. \ref{eq:phase_corr_mg} to the phase of \textsc{IMRPhenomXAS} and inject a non-accelerating GW170817-like binary in zero-noise by setting $A_0 = 2 \times 10^{-43}\, {\rm eV}^{2}$ and then perform the PE to infer LOSA and other parameters using \textsc{IMRPhenomXAS} as the base waveform.

\begin{figure*}
    \centering
    \includegraphics[width=1\linewidth]{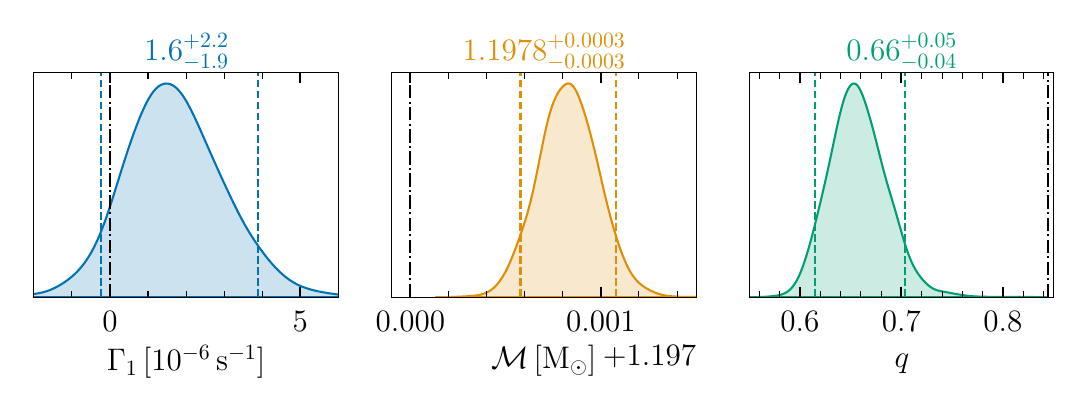}
    \caption{$1d$ posteriors of LOSA (\textit{left panel}), $\mathcal{M}$ (\textit{central panel}), and $q$ (\textit{right panel}) with \textsc{IMRPhenomXAS} as the base waveforms for a non-spinning and non-accelerating GW170817-like CBC at $100 \, {\rm Mpc}$, injected with \textsc{IMRPhenomXAS} together with the modifications due to a finite graviton mass.}
    \label{fig:post_mg}
\end{figure*}

We find the SNR $\rho_{\rm mf}^N$ to be $16.01$. Figure \ref{fig:post_mg} shows the posteriors of $\Gamma_1$, $\mathcal{M}$, and $q$. Even though $\Gamma_1$ is recovered near the $90\%$ CI boundary, $\mathcal{M}$ and $q$ both are completely biased. Hence, a finite mass of graviton can lead to a misinterpretation of a non-accelerating binary as an accelerating one. 

\subsection{Non-acceleration effects near SMBHs}

A CBC that is near an SMBH could have other potential impacts on the waveform beyond just acceleration. We enumerate some of these below and discuss whether they can be confused with a LOSA.

\begin{enumerate}
    \item \textit{Time-varying gravitational redshift}: For CBCs in a circular (outer) orbit around an SMBH, the gravitational redshift will be constant and hence can not create a LOSA-like signature. But for CBCs in a non-circular outer orbit, it will be time-varying and hence can create a LOSA-like signature.
    
    The equivalent of $\Gamma_1 = \dot{z}$ is given by $\dot{z} = -\frac{1}{2} (1 - R_{\rm s} / r)^{-3/2} \left(R_{\rm s}/r^2\right) \dot{r}$, where $R_{\rm s}$ is the Schwarzschild radius of the SMBH, $r$ is the distance of the CBC from the SMBH, and $\dot{r}$ is the radial velocity of the CBC. We find that even for a radial velocity of $3000\, \rm km/s$, the redshift time-variation is one order of magnitude suppressed as compared to the time-variation of the orbit itself.
    
    We further explore the CBC to be in an eccentric outer orbit around an SMBH. We find that, in the case of the most eccentric allowed stable orbit and most extreme values of the true anomaly, variation of gravitational redshift dominates for $a_{\rm s}/R_{\rm s} < 10$, where $a_{\rm s}$ is the semi-major axis of the outer orbit, while the variation of the Doppler shift due to LOSA and the gravitational redshifts match at higher values of $a_{\rm s}/R_{\rm s}$. For other, less extreme cases, the variation of the Doppler shift due to LOSA dominates.
    
    \item \textit{Shapiro delay}: This is a $4 \rm PN$ effect, and  will be relevant for LVK sources if $r \leq R_{\odot}$ for SMBHs, where $R_{\odot}$ is the solar radius \cite{2017ApJ...834..200M}. We are sensitive to LOSA well beyond this radius, where Shapiro delay is suppressed.
    
    \item \textit{Time-dependent lensing magnifications}: Refs. \cite{2020PhRvD.101h3031D} and \cite{ 2022MNRAS.515.3299G} look at the scenarios of time-dependent lensing magnifications when a CBC is near an SMBH. Ref. \cite{ 2022MNRAS.515.3299G} specifically calculates the time-dependent magnifications in the weak and strong lensing regimes for a stellar mass CBC around an SMBH. For the signal durations that we consider in ground-based detectors, the magnifications in the strong lensing regime would just be constant and would just rescale the distance to the object. Therefore, these effects will not lead to the biased inference of LOSA. However, the weak lensing (microlensing) effects could lead to biases as they will modulate the GW waveform.
    
    \item \textit{Repeated signals due to lensing}: These will manifest in the detector as signals separated by $\sim 1-10 \, {\rm s}$ time delays, essentially appearing as overlapping signals in the detector. If the delay is more than the autocorrelation length of the signal (typically $\sim 1 \, {\rm s}$, one can tell apart the two overlapping signals. Even if this condition is not satisfied, overlapping signals will not be degenerate with LOSA---we have checked this explicitly by performing match calculations of non-accelerating but overlapping signals with a free LOSA parameter, finding that the match value peaks at zero LOSA regardless of the injected lensing time delay. 

    \item \textit{Induced eccentricity due to Lidov-Kozai oscillations}: Perturbation by the massive BH can lead to an increase in eccentricities of the inner orbit. References~\cite{Yu:2020dlm, Fang:2019mui} showed that for inner orbits $\lesssim10^{-3}$ AU, this effect is not only subdominant to a LOSA but also to de Sitter precession of the orbital plane of the inner binary. For stellar-mass binaries, the source enters the LVK detection band at a separation $\lesssim 10^{-5}$ AU, the impact of induced eccentricities on the GW phase would be negligible compared to LOSA. However, these oscillations will be detectable in low-frequency detectors and must be accounted for in the analysis.\

\end{enumerate}

\section{The Case of GW190814}\label{sec:GW190814_like_analysis}
In this section, we discuss the curious case of GW190814, which was a GW event detected on August 14, 2019, by the LIGO (Hanford and Livingston) and Virgo detectors \cite{Abbott_2020_GW190814} during their third observing run. It was a merger of a BH of mass ranging between $22.2 - 24.3 \, {\rm M}_{\odot}$ and a compact object of mass ranging between $2.50 - 2.67 \, {\rm M}_{\odot}$. Due to its extreme mass ratio ranging between $0.103 - 0.120$, \cite{2024arXiv240101743H} found it to be a promising merger event that may have happened in an active galactic nucleus (AGN) disk and reported the presence of a third body near the merger based on the LOSA inference. 
However, due to the asymmetric nature of the binary, the GW higher modes were significantly excited, especially the $(l,m) = (3,3)$ mode \cite{Abbott_2020_GW190814}. In addition, marginalization techniques such as phase marginalization can not be used with waveforms having higher modes~\cite{2019PASA...36...10T}. Since \cite{2024arXiv240101743H} only uses the 22-mode leading order LOSA phasing correction and applies it to \textsc{IMRPhenomXPHM} with marginalized likelihoods, the inferred LOSA could be a result of waveform systematics. We investigate biases incurred due to the use of a phase marginalized likelihood to speed up PE in \textsc{Bilby}.

We inject a GW190814-like non-accelerating binary using \textsc{IMRPhenomXHM} in zero noise. We set the signal duration to 32 seconds and injection parameters to: $\{ \mathcal{M} = 6.11 \, {\rm M}_{\odot},\, q = 0.1116,\, D_{\rm L} = 230 \, {\rm Mpc},\, t_{\rm c} = {\rm GPS \, time}\, (1249852257.0),\, \chi_1 = 0.07,\, \chi_2 = 0.02,\, {\rm RA} = 0.24,\, {\rm DEC} = -0.6,\, \theta_{\rm JN} = 0.9,\, \psi = 0.54,\, \phi = 2.95\}$. We then perform PE with \textsc{IMRPhenomXAS} and \textsc{IMRPhenomXHM} as the base waveforms using $\mathcal{U}(5.5, 6.75)$ prior on $\mathcal{M}$, $\mathcal{U}(0.01, 0.5)$ prior on $q$, and ${\rm PL}(10, 2000) \propto D_{\rm L}^2$ prior on $D_{\rm L}$. We fix the geocentric time $t_{\rm c}$ to a GPS time of $1249852257.0$ and keep the priors on the rest of the parameters the same as in Table \ref{tab:priors}. We perform PE using phase marginalized likelihood for both waveforms together with the one without using phase marginalized likelihood for \textsc{IMRPhenomXHM}. Note that we do not use \textsc{Relative Binning} for the \textsc{IMRPhenomXHM} recovery to ensure that no related potential biases are incurred.

\begin{table}[h]
    \centering
    \begin{tabular}{|c|c|c|}
    \hline
         \textbf{Base Waveform} & \textbf{Phase Marginalization} & $\rho^N_{\rm mf}$ \\
         \hline
         \hline
         \textsc{IMRPhenomXAS} & On & 41.07 \\
         \hline
         \textsc{IMRPhenomXHM} & On & 39.75 \\
         \hline
         \textsc{IMRPhenomXHM} & Off & 42.56 \\
         \hline
    \end{tabular}
    \caption{SNRs of GW190814-like non-accelerating injections.}
    \label{tab: snr_190814}
\end{table}

\begin{figure}
    \centering
    \includegraphics[width=1\linewidth]{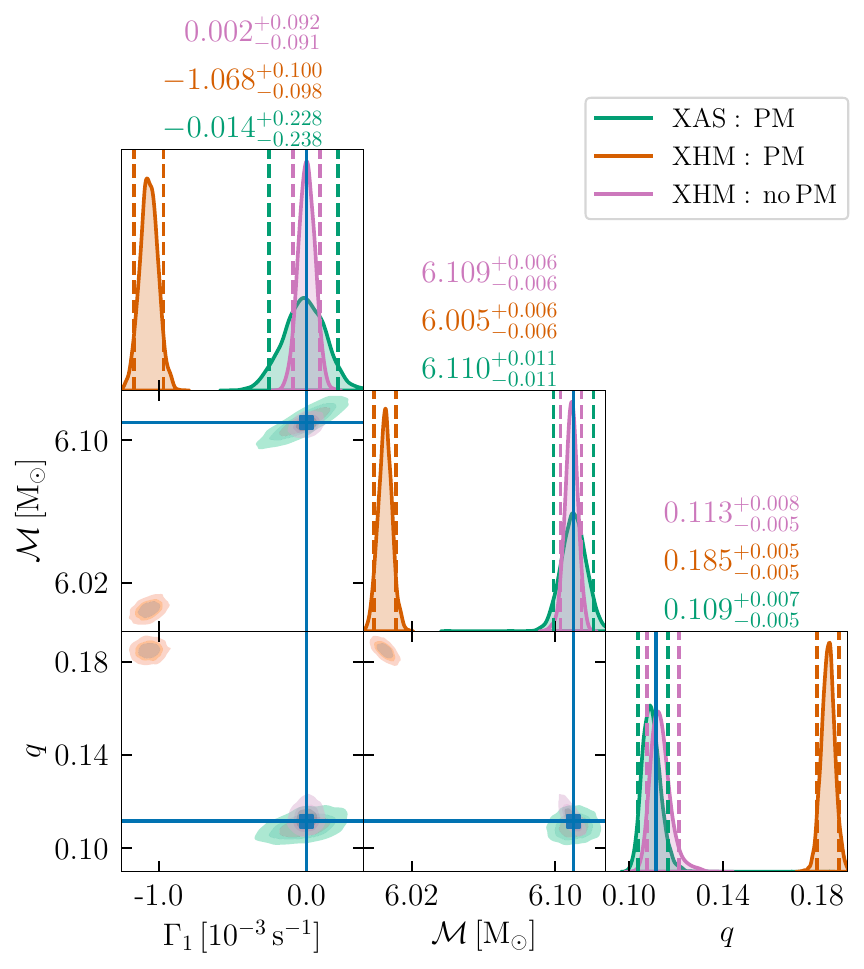}
    \caption{The corner plots of $\Gamma_1$, $\mathcal{M}$, and $q$ highlighting the waveform systematics and effect of phase marginalization for a non-accelerating GW190814-like CBC, injected with \textsc{IMRPhenomXHM} in zero-noise. The green color corresponds to the posteriors recovered with \textsc{IMRPhenomXAS} as the base waveform using the phase marginalization, brown corresponds to the same with \textsc{IMRPhenomXHM} using the phase marginalization, while the magenta corresponds to the same with \textsc{IMRPhenomXHM} without using phase marginalization.} 
    \label{fig:corner_plot_comp_GW190814}
\end{figure}

Figure \ref{fig:corner_plot_comp_GW190814} shows the corner plots of $\Gamma_1$, $\mathcal{M}$, and $q$ for the cases considered above and Table~\ref{tab: snr_190814} shows the corresponding SNRs. Though all parameters are well recovered while using \textsc{IMRPhenomXAS} with phase marginalization and \textsc{IMRPhenomXHM} without phase marginalization, using \textsc{IMRPhenomXHM} with phase marginalization for the recovery results in biased posteriors. Indeed, none of the parameters are recovered using \textsc{IMRPhenomXHM} with phase marginalization --- the setup similar to the one used in \cite{2024arXiv240101743H}. 
We therefore recommend that the results of LOSA inference in \cite{2024arXiv240101743H} must be interpreted with caution. 

\section{The Selection Criteria}\label{sec:sel-crit}
Although Fisher Matrix results of Section \ref{sec:approx_validity} give us a rough idea of the detector frame component masses for which we will get the most robust results of LOSA inference, we also need to consider the fact that LOSA can be mimicked by several GR and non-GR effects, some of which we have already discussed in this paper. Therefore, we consider the injection studies of this paper and present a set of selection criteria for the real GW events observed by the LVK for which we can trust the results of the LOSA inference:
\begin{itemize}
    \item The median of the detector frame total mass $M$ of the CBC as estimated from the PE analyses should be $\leq 10 \, {\rm M}_{\odot}$. LOSA inference should be made on the events passing this criteria, and then the quantity $\vert \Gamma_1 \vert \times t_{\rm sd}$ should be estimated assuming the frequency range used for the analysis and posterior of the chirp mass to make sure it is $\leq 0.01$ at the 90\% CI.
    \item For BNSs, the waveform having tidal corrections should be used.
    \item For signals with significant higher modes, the LOSA analysis should not be carried out by adding the phase corrections to a waveform with higher modes, such as \textsc{IMRPhenomXPHM} because the phase corrections in this paper have been derived only for the $(l, m) = (2, 2)$ mode. Instead, a waveform without higher modes, such as \textsc{IMRPhenomXP} should be used unless the median value of mass ratio $q < 0.25$\footnote{Note, however, that this is a conservative number as we show in the appendix \ref{app_subsec:low_q_hm_inj} that \textsc{IMRPhenomXP} can still be used as a base waveform for $q \geq 0.15$, especially when the SNRs are high.}.
    \item For signals with precessing spins, the LOSA study should be performed only using a waveform with precessing spin\footnote{Note that in precessing-spin scenarios, the aligned spin components generally evolve with frequency, whereas in our implementation (Equation~\ref{eq: phase_corr_AS}), they are held constant and evaluated at the reference frequency. However, we have found no appreciable biases in the recovery for $\chi_p \leq 0.4$.}, such as \textsc{IMRPhenomXP}, unless the median value of $\chi_{\rm p} > 0.4$.
\end{itemize}

Note that we do not see any biases in the LOSA inference (see the Appendix \ref{sec: appendix} for more details) for signals with precessing spins or having higher modes for the scenarios mentioned in the last two points till SNR $> 50$. However, for SNRs $< 25$, we may get biased results. We therefore recommend that if any of our selection criteria are violated, multiple physical models should be considered, and a Bayesian model selection should be performed to determine the model that best fits the data. 

\section{SUMMARY AND DISCUSSION}\label{sec:conclusion}
In this work, we derived the contribution of tidal effects and aligned spins to the phase correction due to LOSA in the GW waveform. We mapped out the parameter space where the approximation Eq. \ref{eq: redshifted_mass} is valid for current generation GW detectors operating at O4 sensitivity. We performed basic sanity checks to ensure that the pipeline was working as expected and explored some physical effects that mimic LOSA. Additionally, we discussed the curious case of GW190814, where we showed that using phase marginalized likelihood in combination with higher modes can produce spurious non-LOSA values even when the source is not accelerating. Finally, we presented a set of selection criteria for the GW events detected by LVK onto which we can apply the LOSA analysis. 

We found that adding the 22-mode phase corrections due to LOSA onto waveforms having higher modes or precessing spins can lead to biased results. However, one can still safely use waveforms without higher modes for analysing a GW signal with higher modes at least up to a mass ratio $\geq 0.25$. Additionally, even though using a waveform with aligned spins as the base waveform for analysing a precessing signal leads to a biased result, one can still safely add the phase corrections derived in this paper for an aligned-spinning system onto a precessing waveform for analysing a precessing signal without biasing the results unless $\chi_{\rm p} > 0.4$. 

We further found that not accounting for eccentricity or tidal effects can lead to a biased inference of LOSA. Given that we have the contribution of the tidal effects to the phase corrections due to LOSA, one can safely infer LOSA for BNSs, but one should be careful of the presence of eccentricity in the signal. Inspired by the fact that eccentricity is a < -3 \rm PN effect, we also explored a couple of non-GR effects namely the modifications due to dipole radiation and a finite mass of graviton that give rise to a $-1{\rm PN}$ and $1 \rm PN$ corrections, respectively, to the phase of GW waveforms. We found that both effects can also lead to biased inferences of LOSA and other parameters describing the observed signal. 

We expect to expand on the parameter space over which our analysis can be reliably used, and also perform additional numerical runs to refine the range of validity of the LOSA-corrected waveforms. We plan to evaluate phase corrections due to LOSA, pertaining to the 33-mode and 44-mode for non-spinning and aligned spin waveforms. With these in place, it should be possible to re-evaluate LOSA for GW190814 to ascertain its host environment. We also plan to evaluate LOSA phase corrections for precessing systems in future work.

\begin{acknowledgments}
    We thank K. G. Arun for his careful reading and feedback. We also thank Nathan Johnson-McDaniel, Michalis Agathos, and Ofek Birnholtz for useful discussions. We gratefully acknowledge computational resources provided by the LIGO Laboratory, a major facility fully funded by the National Science Foundation under Grants PHY-0757058 and PHY-0823459, and the Inter-University Centre for Astronomy \& Astrophysics (IUCAA), Pune, India. AV acknowledges support from the Natural Sciences and Engineering Research Council of Canada (NSERC) (funding reference number 568580). S.G was supported by the Max Planck Society’s Independent Research Group program. S.J.K acknowledges support from Science and Engineering Research Board (SERB) Grants SRG/2023/000419 and MTR/2023/000086.
\end{acknowledgments}

\bibliography{refer}

\section*{Appendix}\label{sec: appendix}

\subsection{Low mass ratio Higher Mode non-accelerating Injections}\label{app_subsec:low_q_hm_inj}
To see how reliably \textsc{IMRPhenomXAS} can be used as base waveform when higher modes are present, we make a couple of non-accelerating CBC injections in zero-noise using \textsc{IMRPhenomXHM} as the base waveform by setting $M = 10 \, {\rm M}_{\odot}$ and $q = 0.15$. For both non-accelerating systems, we consider two luminosity distances: $37 \, {\rm Mpc}$ (high SNR) and $200 \, {\rm Mpc}$ (low SNR). We then recover the paramters using \textsc{IMRPhenomXAS} as the base waveform using $\mathcal{U}(3.2, 3.5)$ prior on $\mathcal{M}$, $\mathcal{U}(0.1, 1)$ prior on $q$, $\mathcal{U}(0, 0.5)$ prior on $\chi_{1,2}$, and ${\rm PL}(10, 1000) \propto D_{\rm L}^2$ prior on $D_{\rm L}$.

\begin{figure}[h]
    \centering
    \includegraphics[width=1\linewidth]{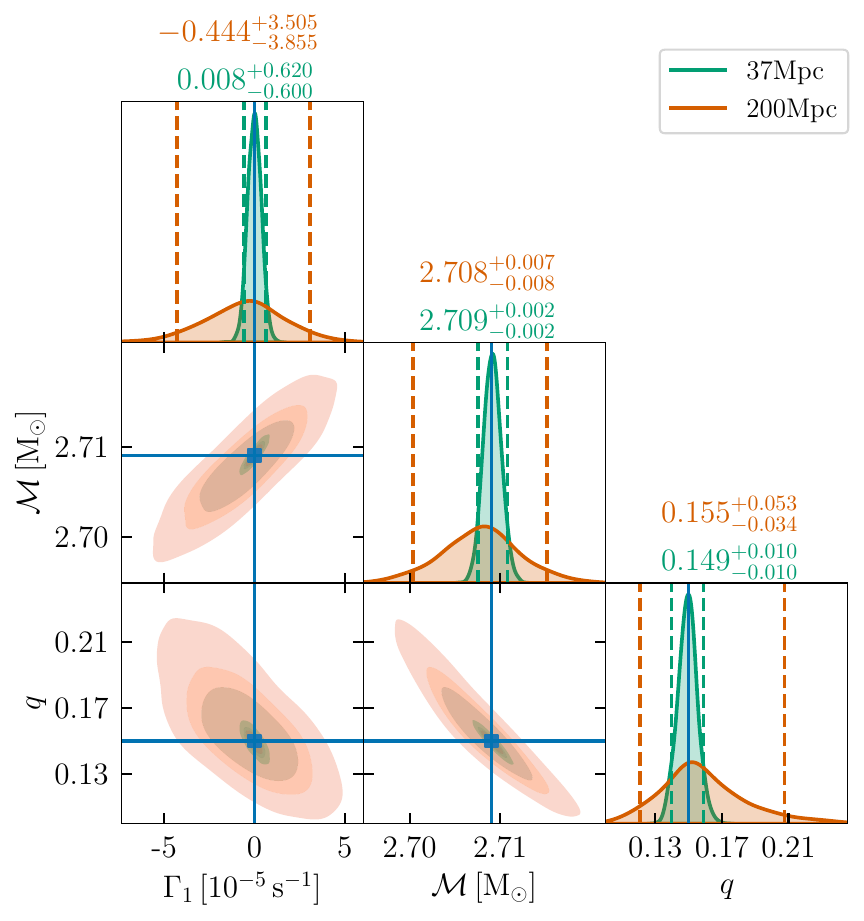}
    \caption{The corner plots of $\Gamma_1$, $\mathcal{M}$, and $q$ recovered using \textsc{IMRPhenomXAS} as the base waveform for the non-accelerating \textsc{IMRPhenomXHM} injections of CBC of total mass $M = 10 \, {\rm M}_{\odot}$ and mass ratio $q = 0.15$ with $D_{\rm L}$s: $37 \, {\rm Mpc}$ (green) and $200 \, {\rm Mpc}$ (brown).}
    \label{fig:app_low_q_hm_inj}
\end{figure}

For non-accelerating injections, we find the SNRs $\rho_{\rm mf}^N$ to be $94.38$ for $37 \, {\rm Mpc}$ injection and $17.26$ for $200 \, {\rm Mpc}$. Figure \ref{fig:app_low_q_hm_inj} shows the corner plots of $\Gamma_1$, $\mathcal{M}$, and $q$. We find that all parameters are recovered within the $90\%$ credible interval. Even though zero $\Gamma_1$ is recovered in both injections, the median values are not zero. Hence, any inference about the environment of a CBC based on LOSA inference should be made with caution unless zero $\Gamma_1$ is completely ruled out. 

\subsection{Low SNR Precessing Injections}\label{app_subsec:low_snr_prec_inj}
In Section~\ref{subsec:Prec_as_LM}, we considered very high SNR injections. To see how low SNRs affect the recovery of parameters after adding the phase correction due to LOSA having AS contribution on \textsc{IMRPhenomXP} with
precessing spins as the base waveform, we re-injected the binary considered in Section \ref{subsec:Prec_as_LM} with different luminosity distances ($200 \, {\rm Mpc}$ and $100 \, {\rm Mpc}$) in zero-noise using \textsc{SEOBNRv5PHM} as the base waveform. 

\begin{figure}[h]
    \centering
    \includegraphics[width=1\linewidth]{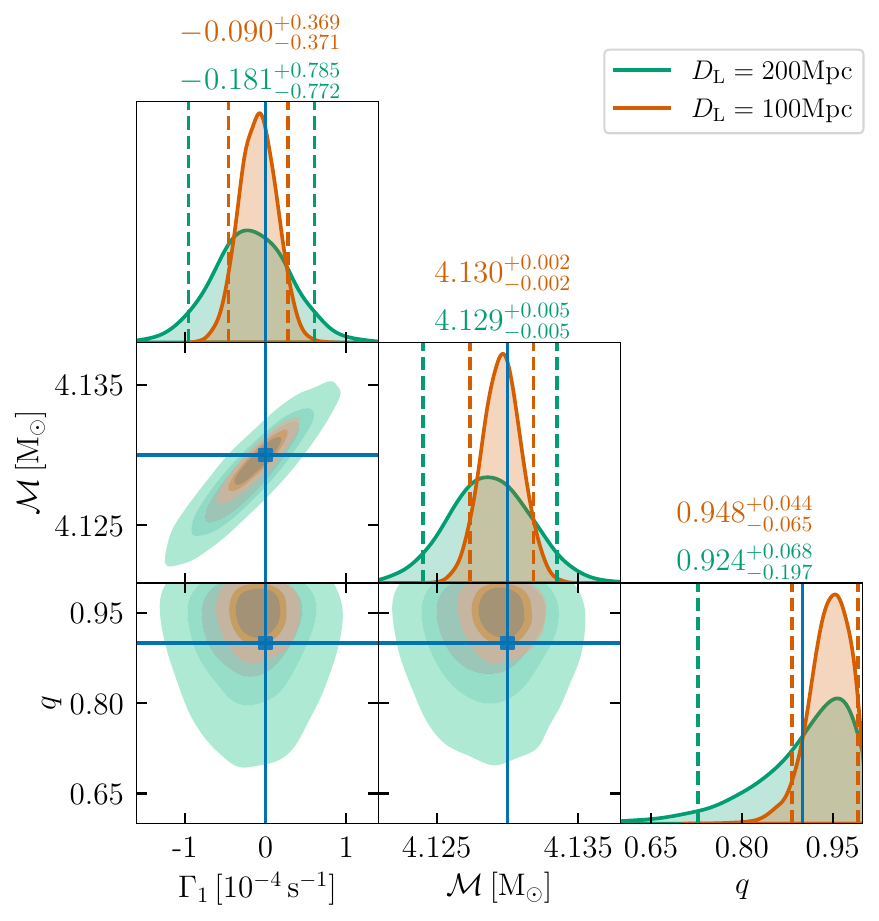}
    \caption{The corner plots of $\Gamma_1$, $\mathcal{M}$, and $q$ recovered using \textsc{IMRPhenomXP} with precessing spins as the base waveform for the injection used in subsection \ref{subsec:Prec_as_LM} with different $D_{\rm L}$s: $200 \, {\rm Mpc}$ (green) and $100 \, {\rm Mpc}$ (brown).}
    \label{fig:app_low_snr_prec_inj}
\end{figure}

For the $200 \, {\rm Mpc}$ injection, we find the SNR $\rho^N_{\rm mf}$ to be $26.13$, while for the $100 \, {\rm Mpc}$ injection we find the same number to be $52.69$. Figure \ref{fig:app_low_snr_prec_inj} shows the corner plots of $\Gamma_1$, $\mathcal{M}$, and $q$ recovered using \textsc{IMRPhenomXP} with precessing spins as the base waveform for these injections. We find that $\mathcal{M}$ and $\Gamma_1$ are very well recovered in both cases, but $q$ is recovered near the boundary for the $100\, {\rm Mpc}$ injection. Though it may seem strange by only looking at the posterior of $q$ that it is well recovered for the lowest SNR case considered here, i.e., the $200\, \rm Mpc$ injection, it is not. This is an artifact of the posterior of $q$ broadening due to low SNR. 

\end{document}

%% file: preamble.tex
\usepackage[T1]{fontenc}
\usepackage[utf8]{inputenc}
\DeclareUnicodeCharacter{2009}{\nobreakspace}
\usepackage{lmodern}
\usepackage{times}
\usepackage[varg]{txfonts}
\usepackage[normalem]{ulem}
\usepackage{verbatim}
\usepackage[dvipsnames, usenames]{xcolor}
\definecolor{linkcolor}{rgb}{0.0,0.3,0.5}
\usepackage[hypertexnames=false, unicode, colorlinks=true, linkcolor=linkcolor,
citecolor=linkcolor, filecolor=linkcolor,urlcolor=linkcolor,
pdfusetitle]{hyperref}
\usepackage[all]{hypcap}
\usepackage{graphicx}
\usepackage{xspace}
\usepackage{amssymb}
\usepackage{bm} 
\usepackage{microtype}
\usepackage[english]{babel}
\usepackage{aas_macros}